# Developing for personalised learning: the long road from educational objectives to development and feedback

George Tsatiris, Kostas Karpouzis
Artificial Intelligence and Learning Systems Lab
National Technical University of Athens

gtsatiris@image.ntua.gr, kkarpou@cs.ntua.gr

This paper describes the design and implementation work done to support the functional and teaching requirements of iRead [1], a 4-year EU-funded project on personalised learning technologies to support reading skills. The core software applications developed in the project are a reader application, which highlights specific parts of the words contained in the text, given specific criteria, and a serious game, which consists of a series of gamified activities utilising words and sentences. The foundation of these applications and the software infrastructure that provides access to the content consists of language models for English, German, Spanish and Greek (for primary school children), and English as a foreign language, including children with dyslexia; following the definition of extensive phonological and syntactic models for these languages, the linguists in the project worked together with teachers to define the learning objectives for each of the target age groups, as well as the sequence in which each language feature should be taught. The sequencing of these features, including which prerequisites should be taught and mastered by the students before moving on to more advanced features, was encoded in a tree-like hierarchical graph; essentially, this graph encapsulates both the language model (i.e. the features that make up each word or sentence, at least at the given language level) and the teaching model, represented by the selection of necessary features for each school year and the succession in which they should be taught. When a new student registers with the iRead system, this graph is instantiated as a user profile, with different values of mastery for each feature, depending on the students' age.

This is where the adaptivity component in iRead kicks in, first by utilising the mastery levels for each feature to select proper content from the project resource engine (dictionaries and texts) and then by updating the student's model based on their performance in each language game they play [2]; when the mastery level for a given feature surpasses a selected threshold (75%), subsequent features in the model hierarchy become available to play with, provided that all prerequisites for them have been met. In the context of iRead, the game content consists of selecting a particular game activity, a language feature to work with, and a set of words or a sentence that corresponds to that feature (e.g. a particular letter, phoneme or a sequence of phonemes). The content selection process starts with the given student's model, i.e. the mastery level for each open feature; then it selects the content for each session by filtering the available resources with a set of rules defined by the project researchers after productive consultation sessions with the teachers collaborating with them. These rules were first defined in verbal form, in order to promote the teaching objectives of the games, with each of them corresponding to a particular pedagogical rationale. For instance, when multiple features are open (available to play with), the Adaptation component sorts them by taking into account how many times each of the features has been used in earlier games and how well the student has previously performed when presented with that feature. The reasoning here is that students should start from an easier feature and should not be playing a feature they have not done well in recently, thus fostering motivation and efficacy. Other rules combine the feature mastery level achieved in previous games with the number of gameplay sessions since that feature was last used to reinforce learning, by reopening a fully mastered feature after ten games have been played since it was last used. The assumption here is that the student has fully mastered that feature, but they must repeat it once in a while, to showcase their progress. Finally, a number of feature selection rules deal with students not progressing as expected or not having truly mastered the language features which correspond to their age level: if a feature has been practised twice and the feature mastery is not improving, the mastery level for that feature and its prerequisites is reduced, so that both can be revisited in future sessions. This content selection strategy treats the issue that the iRead system assumes that students in a given age group have already mastered certain language features and allows them to go back to required knowledge, if there is no system evidence that it has been acquired. In addition to selecting proper word content, the iRead adaptivity system utilises a rule-based strategy to select specific game activities to utilise those words: if a feature has not been previously used in a game for the particular student, then the game selected should promote accuracy in using that language characteristic, before moving on to games which stimulate automaticity.

The second part of the adaptation component in iRead has to do with re-evaluating the value of mastery for the language feature used in a game. During the consultation sessions, teachers mentioned a number of requirements for this process: changes in mastery values should not be abrupt, especially when students make an occasional error in one of the activities, and at the same time, help students demonstrate complete mastery of a feature within a handful of gaming sessions, allowing them to move on to more advanced and interesting features. A mathematical definition that accommodates these requirements, while leaving room for experimentation and adjustments of the process, is that of Exponential Moving Average (EMA) [3]: essentially, this takes into account previous attempts at a particular feature (previous game sessions) but gives more weight to recent attempts. The number of previous attempts to consider may be defined by each implementation; in iRead, we implemented the complete definition, but chose to consider only the previous value of mastery, when calculating the next one. This averaging process allows students to show complete mastery in just three games, since each newly opened feature is initialised with a value of 5 and after three perfect games reaches the maximum value of 10. In addition, in case the student makes one or more errors during game play, the respective mastery value may be reduced by 1 at maximum. This mechanic, along with the rules which prioritise features given the recent gameplay attempts, allows students to practise different language content, without being stuck with difficult features.

Early evaluation has shown that the automated content selection and the profile re-evaluation process are quite close to what teachers expect and provides suitable and interesting content for the students. Even though the re-evaluation mechanic allows unlocking subsequent features quite easily, there have been reports that students are being given the same features to play with during numerous successive game sessions. However, after going through the system logs of gameplay results and mastery evaluations, we reached the conclusion that this reflects the design of the respective language model, which imposed a large number of prerequisites to be unlocked before moving on to more complex features. This effectively illustrates the interplay between the different iRead components: the features which describe each language model, the graph of prerequisites which describes the sequencing embedded in the learning process, the mastery levels for each feature which reflect student performance, and the adaptation and re-evaluation rules described above, which prioritise content to implement teaching objectives.

The large-scale evaluation phase in schools across Europe is already underway, with more than 2000 students taking part. Even though it has been disrupted by the pandemic and schools closing down, we expect gameplay logs to keep coming in from students playing the games at home. Processing these logs will allow us to revisit specific parts of the adaptation component, primarily the content selection rules and the mastery re-evaluation implementation.